\documentclass[11pt,twoside]{article}
\usepackage{./asp2014}

\aspSuppressVolSlug
\resetcounters

\begin{document}

\title{Explorations of Dusty Debris Disk Geometry}
\author{Dennihy, E.,$^1$ Debes, John H.,$^2$ Clemens, J. C$^1$ \\
\affil{$^1$Physics and Astronomy Department, University of North Carolina at Chapel Hill, Chapel Hill, NC 27599; \email{ edennihy@unc.edu, clemens@unc.edu}}
\affil{$^2$Space Telescope Science Institute, Baltimore, MD 21218; \email{debes@stsci.edu}}}

\paperauthor{Dennihy, E.}{edennihy@unc.edu}{}{University of North Carolina}{Department of Physics and Astronomy}{Chapel Hill}{NC}{27599}{United States}
\paperauthor{Debes, John H.}{debes@stsci.edu}{}{Space Telescope Science Institute}{}{Baltimore}{MD}{21218}{United States}
\paperauthor{Clemens, J. C.}{clemens@unc.edu}{}{University of North Carolina}{Department of Physics and Astronomy}{Chapel Hill}{NC}{27599}{United States}

\begin{abstract}
As the sample of white dwarfs with signatures of planetary systems has grown, statistical studies have begun to suggest our picture of compact debris disk formation from disrupted planetary bodies is incomplete. Here we present the results of an effort to extend the preferred dust disk model introduced by \citet{jur03} to include elliptical geometries. We apply this model the observed distribution of fractional infrared luminosities, and explore the difference in preferred parameter spaces for a circular and highly elliptical model on a well-studied dusty white dwarf. 
\end{abstract}

\section{The Current Picture of Dusty Infrared Excesses}
To date, several dozen white dwarfs with compact circumstellar dust disks have been confirmed via the detection of infrared radiation in excess of their stellar photospheres. The observed excess offers the chance to constrain a handful of properties of the dust, including rough orbital parameters. The favored model for the origin of the dust was introduced by \citet{jur03}, wherein an asteroid on an initially highly eccentric orbit is tidally shredded, and the resulting dust settles into a compact optically thick and vertically thin disk. The orbits of the dust resulting from the tidal disruption are then expected to be constrained within two physical boundaries: the outer edge should be broadly consistent with the tidal disruption radius, beyond which asteroids are expected to survive their flyby encounters, while the inner edge should be able to extend to the sublimation radius of the dust, within which no dust can survive the intense radiation of the white dwarf.

The \citet{jur03} model has seen great success in it's ability to explain both the abundance patterns seen in the material accreted onto the white dwarf surface, and to reproduce the observed infrared excess. The inner and outer radii of best-fitted models of the infrared excess are broadly consistent with the expected physical boundaries and, despite extensive searches, no evidence for a substantial amount of dust existing beyond the the tidal disruption radius has been found \citep{far16}. 

\section{The Motivation for Extending the Model}
\articlefigure[scale=0.5]{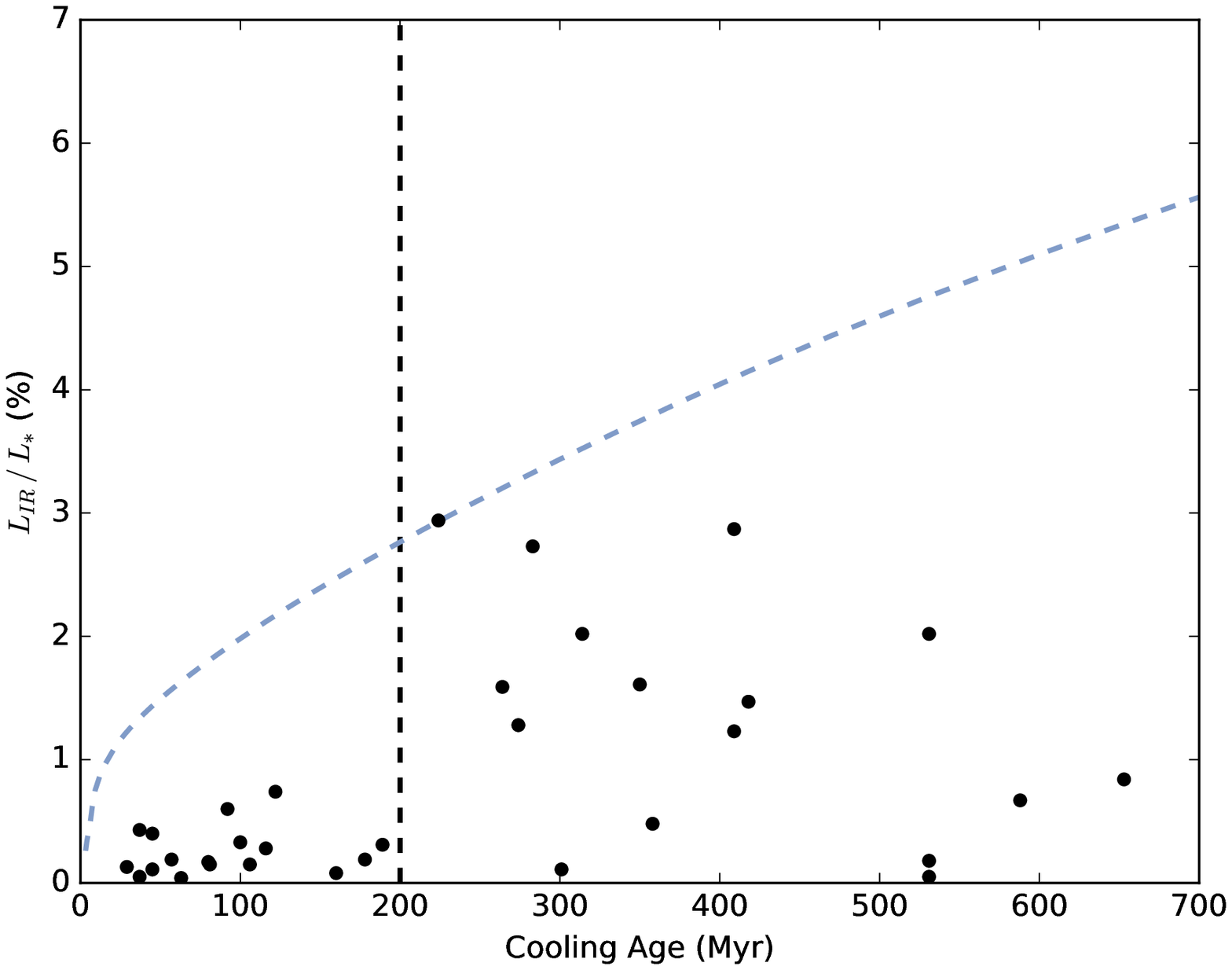}{fig1}{The reproduced central panel of Figure 4 from \citet{roc15}. The dashed blue line tracks the expected fractional infrared luminosity from a face-on disk filling its available orbital space, while the vertical dashed black line delineates the bins of young and old white dwarfs}

The sample of dusty white dwarf systems has now grown to the point to where we can begin examining the properties of the sample as a whole, rather than individual objects. One such study \citep{roc15} examined the distribution of fractional infrared luminosities of all known dust systems as a function of white dwarf cooling age. In Figure 1, we show a reproduced version of the central panel from Figure 4 of \citet{roc15}. Plotted on the vertical axis is the observed fractional infrared luminosity, or a ratio of the brightness of the infrared source to the white dwarf, and on the horizontal axis the white dwarf cooling age. The dashed blue line tracks the fractional infrared luminosity of a disk with dust which fills all of its available orbital space between the sublimation and tidal disruption radius and is seen face on, so the complete disk area is visible to the observer. This line represents a maximum fractional infrared luminosity under a few assumptions of the dust properties. 
	
The observed fractional infrared luminosity is not expected to always lie along the dashed blue line. Instead we expect a smooth distribution of fractional infrared luminosities below the dashed blue line, reflecting a sample with random disk inclinations. While that distribution seems apparent for discs around cooler white dwarfs, there is an abrupt drop in the observed fractional disc luminosity at about 200 Myr, identified by the vertical dashed black line. Given the number of systems in the bin of younger white dwarfs, there is a very low probability that the reduced fractional infrared luminosity (relative to the dashed blue line) is due to high inclination. In other words, it is statistically improbable that all of these disks are being observed at high inclination. Instead, this work clearly suggests the discs around younger white dwarfs are inherently fainter.

One way to make fainter disks without adjusting the inclination, as suggested by \citet{roc15}, is to just assume they are not filling all of their available orbital space. In other words the discs around young white dwarfs could just be narrower than their counterparts around older white dwarfs. These suggestions are supported by \citet{boc11}, which demonstrated that the global evolution of an optically thick dust disk trends towards narrow rings, and the increased radiation of the younger white dwarfs could accelerate this process. Here we explore an alternative origin for the fainter infrared excesses by considering the effects an elliptical dust distribution would have on the observed spectral energy distribution. This work is motivated by the known elliptical origins of the dust, and the relatively unknown timescales for circularization \citep{ver15}. 

\section{The Effects of Elliptical Geometry}
\articlefigure[scale=0.5]{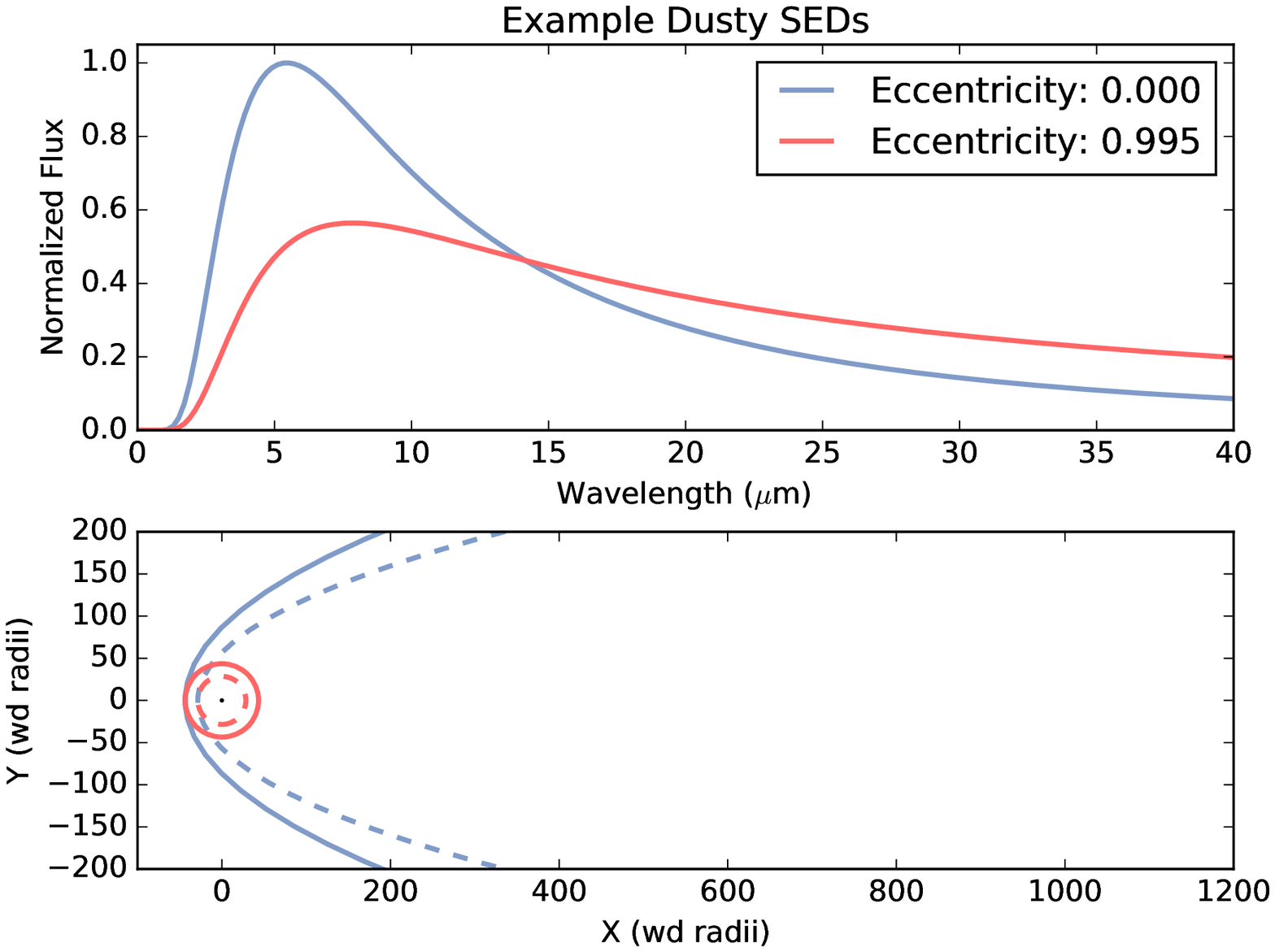}{fig2}{Elliptical and Circular spectral energy distributions with corresponding face on geometrical projections.}

As introduced in \citet{den16}, we begin by adopting the same passive, flat, and opaque assumptions about the dust particles as considered by \cite{jur03}. We consider the white dwarf to be at the center of our radial coordinate system, and the inner and outer ellipses that bound the disk to be described by the parameters $a_{\rm in}$, $a_{\rm out}$, and $e$ where $a_{\rm in}$ and $a_{\rm out}$ define the semi-major axes of the inner and outer ellipses, and $e$ defines the eccentricity of the ellipses which we hold fixed for all nested ellipses. The ellipses are confocally nested with the white dwarf at one focus. In this way, the coordinate $r$ specifies the distance to a dust particle from the white dwarf which uniquely determines the particle temperature under the assumption of an optically thick, flat disk. The inner/outer ellipses which bound the disk are then described by the equation:
\begin{equation}
r_{\rm in/out}(\theta) = \frac{a_{\rm in/out}(1-e^2)}{1-e\cos\theta}
\end{equation} 
and our integral for the observed monochromatic flux as a function of frequency becomes:
\begin{equation}
F_\nu = \frac{\cos i}{D^2}\int_{0}^{2\pi}d\theta \int_{r_{\rm in}(\theta)}^{r_{\rm out}(\theta)} B_\nu\left(T_{\rm ring}(r)\right)rdr
\end{equation}
where $i$ is the inclination of the disk and $D$ is the distance of the system. 

An illustration of this geometry and its effect on the spectral energy distribution is shown in Figure 2. Note how as the eccentricity grows and material is redistributed to larger distances from the white dwarf, the monochromatic flux in the resulting spectral energy distribution is redistributed to longer wavelengths, and the luminosity in bands we typically observe (somewhere between 1-10 microns) is decreased, thereby reducing the observed infrared flux of the disk without adjusting its inclination.

\articlefigure[scale=0.5]{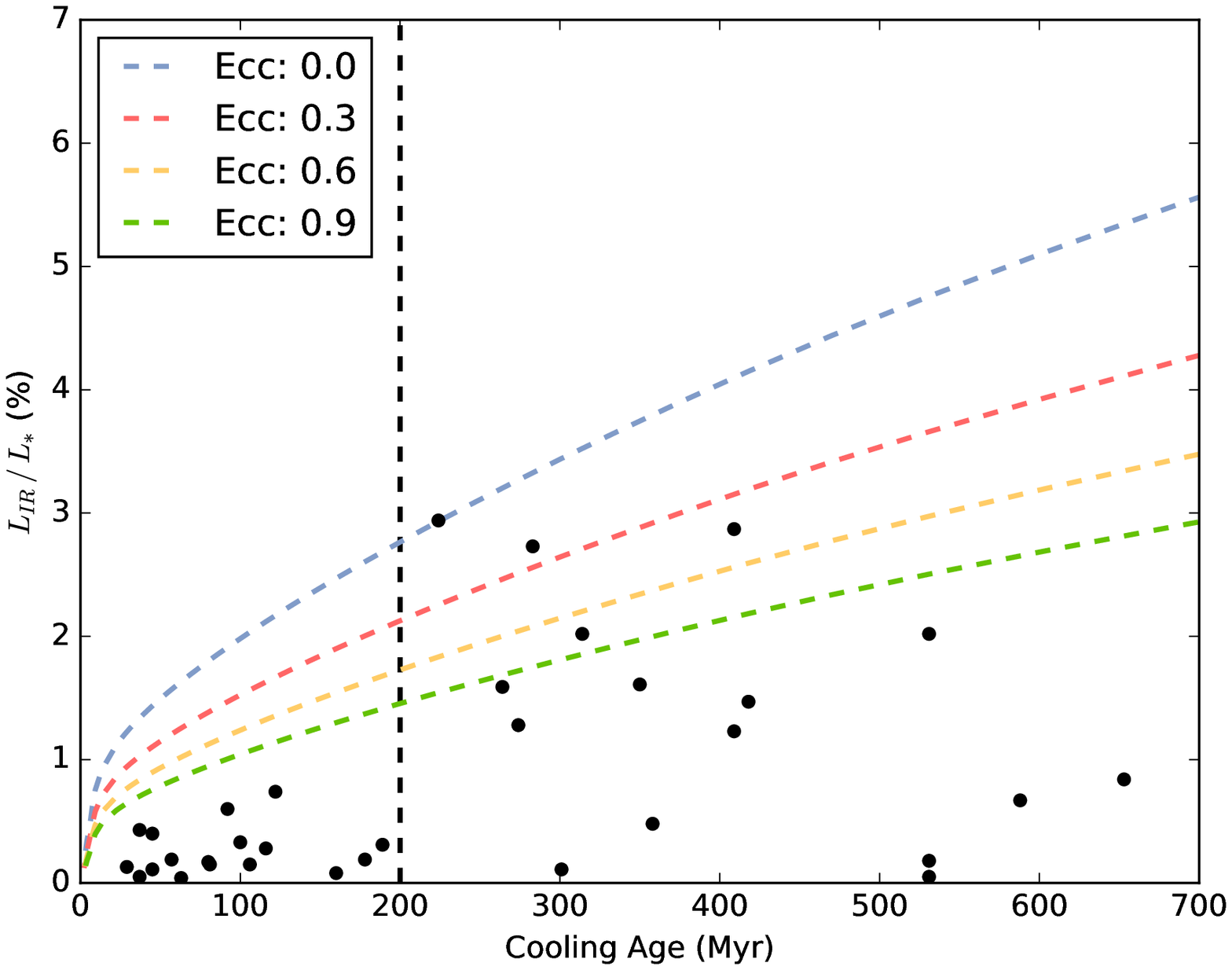}{fig3}{The distribution of fractional infrared luminosities compared with face-on disk models of increasing ellipticity.}

In Figure 3 we re-plot Figure 1 with additional models of increasing eccentricity. As expected, lines of increasing eccentricity have a reduced fractional infrared luminosity, and better approximate the maximum fraction infrared luminosity of the younger white dwarfs under the assumption that the disks fill all of the available orbital space and have randomly observed inclinations. As an alternative to the narrow rings interpretation, we propose that eccentricity could play a role in the difference between the younger and older sample of white dwarfs. Given that we know the dusty material must begin on highly eccentric orbits, the difference between the older and younger samples on the fractional infrared luminosity plot could be explained by the younger disks hosting more frequent disruption events, and therefore having disks of higher mean eccentricity. To gain a deeper insight into how the elliptical geometry effects the inferred dusty properties, we also applied the model to an individual, well-studied young dusty white dwarf as a case study. 

\section{Case Study: SDSSJ0845+2257 (Ton 345)}
For our case study we chose a young white dwarf with a large suite of infrared photometry points including a full complement of spitzer IRAC, IRS 16 micron and MIPS 24 micron data \citep{brin12}. Ton 345 also has independent JHK photometry sources and is known to host a variable gaseous disk \citep{mel12, wil15}. 

In Figure 4 we present the best fit circular and highly eccentric disc models, as well as geometric representations of the disks in a face on configuration. Note the best-fit SEDs were not face-on, but rather at 77 and 59 degrees respectively. So focusing on the best-fitted elliptical disk, it not only fails to fill in more of the available orbital space between the tidal disruption radius and the sublimation radius at peri-center, it actually gets narrower! So at least for this young white dwarf, a narrow disk (relative to the total available space) is preferred independent of eccentricity. But another interesting consequence of these models is that a significant amount of the disk material is now outside of the tidal disruption radius at around 100 white dwarf radii, and yet we are not over-predicting the 24 micron flux. In contrast, circularly distributed dust at this radius would greatly over-predict the 24 micron flux. 

\articlefiguretwo{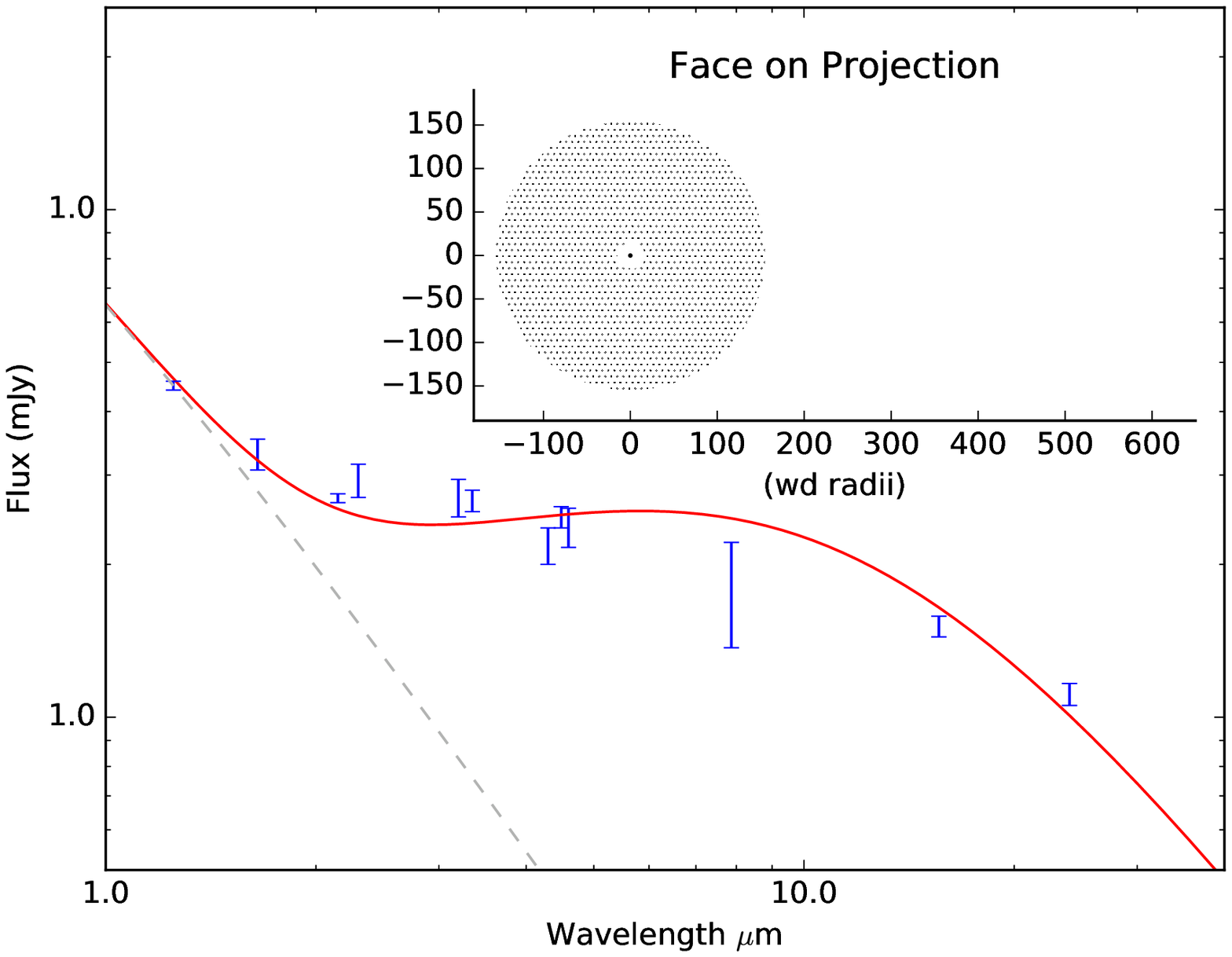}{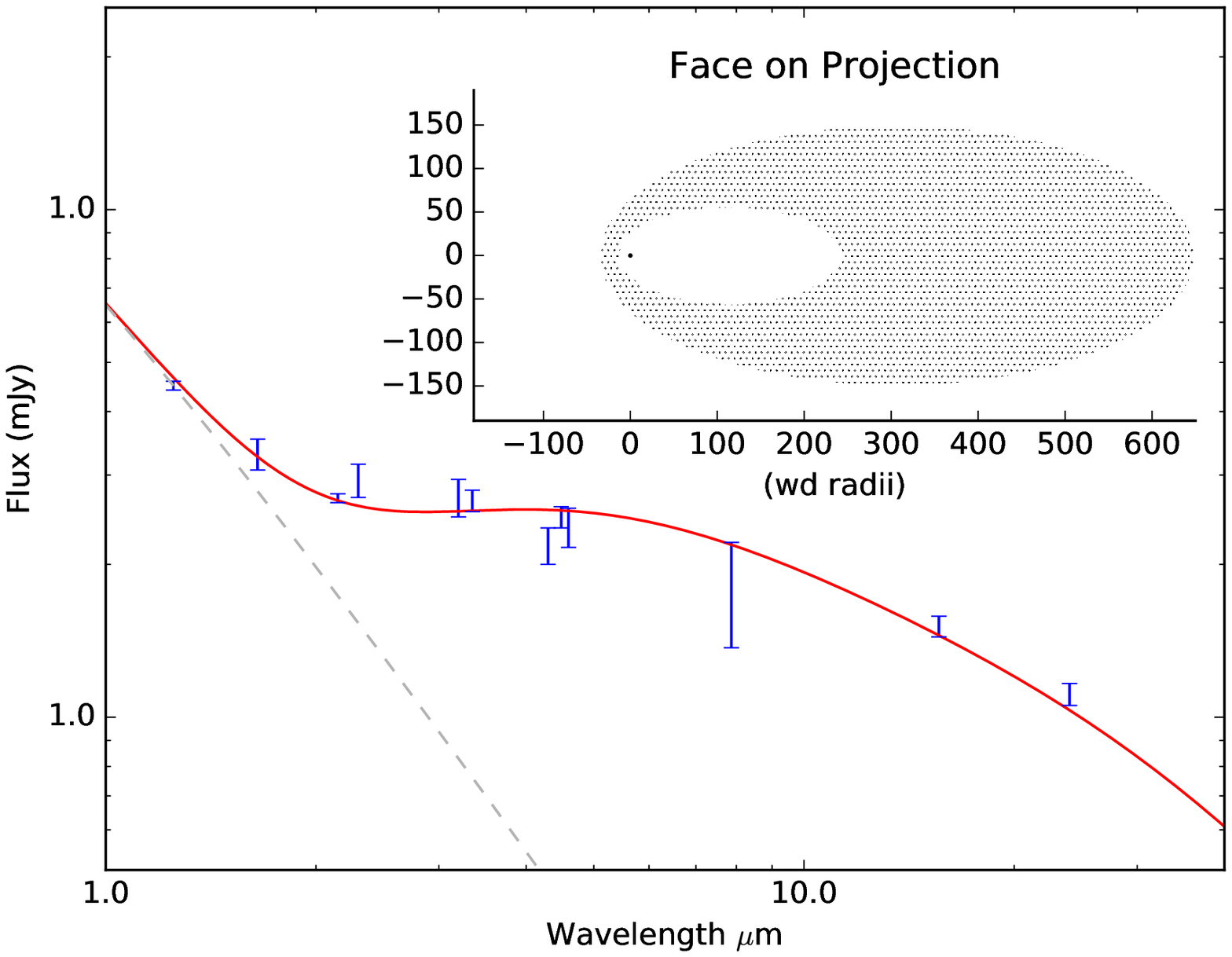}{fig4}{Comparison of best fit solutions to Ton 345 IR excess  \emph{Left:} High inclination circular model  \emph{Right:} Low inclination elliptical model}

\section{Caveats and Future Work}
While it is true that the dust starts off on highly elliptical orbits just after disruption, most of the physical processes we associate with these dust disks seek to circularize material. \citet{ver15} recently demonstrated that when considering the evolution of an individual ring of particles under Poynting-Robertson drag alone, the dust can be expected to circularize in hundreds to tens of thousands of years for the younger white dwarfs in this sample. If the circularization timescale is too short compared to the lifetime of the disk, we would not expect to observe many eccentric disks.

Despite the theoretical concerns, it is not yet clear that the data rule out elliptical geometries. For future work we intend to apply the case-study style analyses to the greater sample of white dwarfs which have 24 micron measurements and upper limits, and determine whether current or future observations can feasibly differentiate between the two geometries. 

\acknowledgements The authors would like to acknowledge B. H. Dunlap, J. T. Fuchs, and J. J. Hermes for useful discussions when developing this work. E. Dennihy and J. C. Clemens acknowledge the support of the National Science Foundation, under award AST-1413001.


\end{document}